\documentclass[aps,pre,preprint,showpacs,groupeaddress,graphics]{revtex4}
\usepackage{graphicx}
\usepackage{dcolumn}
\usepackage{bm}

\begin{document}
\title{Phase transitions in the two-dimensional super-antiferromagnetic Ising 
model with next-nearest-neighbor interactions}
\author{A. \surname{Saguia}}
\email{amen@if.uff.br}
\affiliation{Departamento de F\'{\i}sica, Universidade Federal Fluminense\\
Av. Litor\^anea s/n,  Niter\'oi, 24210-340, RJ, Brazil}
\author{B. \surname{Boechat}}
\email{bmbp@if.uff.br}
\affiliation{Departamento de F\'{\i}sica, Universidade Federal Fluminense\\
Av. Litor\^anea s/n,  Niter\'oi, 24210-340, RJ, Brazil}
\author{O.F. \surname{de Alcantara Bonfim}}
\email{bonfim@up.edu}
\affiliation{Department of Physics, University of Portland, Portland, Oregon 97203, USA}
\author{J. \surname{Florencio}}
\email{jfj@if.uff.br}
\affiliation{Departamento de F\'{\i}sica, Universidade Federal Fluminense\\
Av. Litor\^anea s/n,  Niter\'oi, 24210-340, RJ, Brazil}

\date{\today}

\begin{abstract}

We use Monte Carlo and Transfer Matrix methods
in combination with extrapolation schemes to determine the 
phase diagram of the 2D super-antiferromagnetic (SAF) Ising model
with next-nearest-neighbor ({\em nnn}) interactions in a magnetic field.
The interactions between nearest-neighbor ({\em nn}) spins 
are ferromagnetic along  {\em x}, and antiferromagnetic along 
{\em y}.  We find that for sufficiently low temperatures and fields,
there exists a region limited by a critical line of 2nd-order transitions
separating a SAF phase from a magnetically induced paramagnetic phase.  
We did not find any region with either first-order transition or with 
re-entrant behavior.
The {\em nnn} couplings produce either an expansion or a contraction
of the SAF phase.  Expansion occurs when the
interactions are antiferromagnetic, and contraction when
they are ferromagnetic.  There is a critical ratio $R_c = \frac{1}{2}$ between
{\em nnn}- and {\em nn}-couplings, beyond which the SAF phase no
longer exists.

\end{abstract}


\maketitle

One system that has drawn considerable interest recently is the 
$s = 1/2$ super-antiferromagnetic (SAF) Ising model on a 
square lattice in the presence of a magnetic field~\cite{Net06,Via09,Din09,Que09}.
The model is described by the Ising interactions with a
special kind of anisotropy,  ferromagnetic  $J_x$ along $x$ and
antiferromagnetic $J_y$ along $y$.
In the absence of an external field, the ground-state
consists of alternating rows of up- and down-spins.
Such ordering is known as the SAF order.
Also, Onsager's exact solution for the 2D Ising model applies
here~\cite{Ons44}.  There is a critical temperature $T_c$, which
separates the low-temperature phase with SAF order from
the paramagnetic phase.
In particular, for $J_x = J_y= J_1$, $T_c/J_1= 2/\ln (1+\sqrt 2) \simeq 2.269$.  
At $T = 0$, an applied external magnetic field $H$ destroys
the SAF order at $H_c = 2J_1$,
where all the spins become aligned with the field~\cite{Rot90}.

%%%%%%%%%%      Figure 1      %%%%%%%%%%%%%%
\begin{figure}
\includegraphics[width=6.0cm,angle=0]{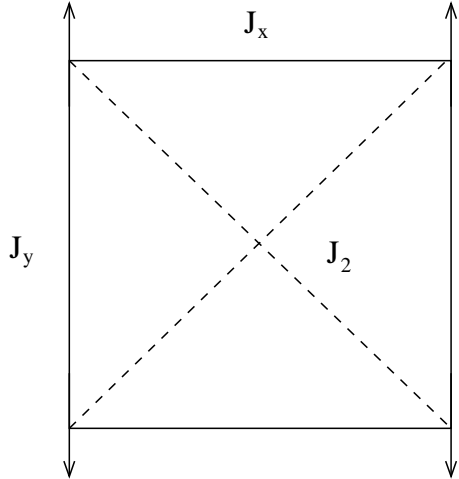}
\caption{ \label{fig:figure1} Energy couplings between
neighboring spins of the SAF Ising model.
}
\end{figure}
%%%%%%%%%%%%%%%%%%%%%%%%%%%%%%%

The phase diagram of the model in the ($H$--$T$) plane
has been studied using different approaches.  One feature
that has caused controversy concerns the re-entrant behavior
in the phase diagram found by 
some authors~\cite{Cha79,Wan97,Net06}.
Such behavior is absent in other 
studies~\cite{Kin75,Rot90,Din09,Via09,Que09}.
The most recent studies in the literature point to the dismissal of
re-entrant behavior.  It seems, however, that more
scrutiny is needed to clear up this controversy.

%%%%%%%%%%      Figure 2      %%%%%%%%%%%%%%
\begin{figure}
\includegraphics[width=8.0cm,angle=0]{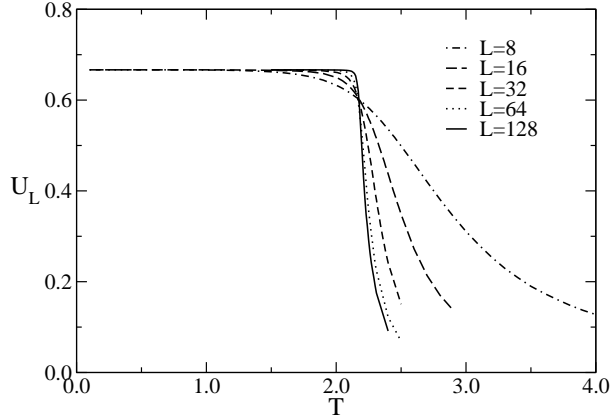}
\caption{ \label{fig:figure2}
Fourth-order cumulant of SAF magnetization {\em versus}
temperature for several lattice sizes $L$, obtained from
Monte Carlo simulations.  The curves
cross nearly at the same temperature.
}
\end{figure}
%%%%%%%%%%%%%%%%%%%%%%%%%%%%%%%

The addition of {\em nnn} interactions may induce
new phases with different orderings and multi-critical points.  
Consider the case of a closely related system, 
the 2D Ising model with antiferromagnetic (AF) interactions.  
In that model,
the phase diagram without {\em nnn} interactions consists of
a second-order critical line separating 
the low-temperature AF phase from 
a paramagnetic phase. 
The inclusion of {\em nnn} ferromagnetic interactions  
reinforces the checkerboard AF order and
causes the system to show
tricritical behavior.    That is characterized by the presence 
of a tricritical point ($H_t,T_t$) in the phase diagram 
line where the transition changes from second to 
first order~\cite{Lan72,Bin80,Her83}. 

The purpose of this work is to investigate the influence  
of  {\em nnn} couplings on the phase transitions
of the 2D SAF Ising model in a uniform magnetic field. 
The Hamiltonian is 
$$
{\cal H} = - J_{x}\sum_{i,j} S^z_{i,j}S^z_{i+1,j}
+ J_{y} \sum_{i,j} S^z_{i,j}S^z_{i,j+1}  
$$
\begin{equation}
- J_{2}\sum_{i,j}(S^z_{i,j}S^z_{i+1,j+1} + S^z_{i+1,j}S^z_{i,j+1})
- H \sum_{i,j} S^z_{i,j},
\label{H}
\end{equation}
where $S^z_i$ can take the values $\pm 1$. 
The parameters $J_x$  and $J_y$ are energy couplings between {\em nn} spins
along $x$ and $y$, respectively.  $J_2$ is the 
coupling between {\em nnn} spins, and $H$
the magnetic field.

In this work we assume $J_x = J_y = J_1 > 0$, whereas
$J_2$ can be either positive or negative.  For simplicity, from here on
we use the notation $R= J_2/J_1$, and set $J_1 = 1$ as the 
energy unit.
Figure~1 shows the energy couplings that appear in Eq.~1.

%%%%%%%%%%      Figure 3      %%%%%%%%%%%%%%
\begin{figure}
\includegraphics[width=8.0cm,angle=0]{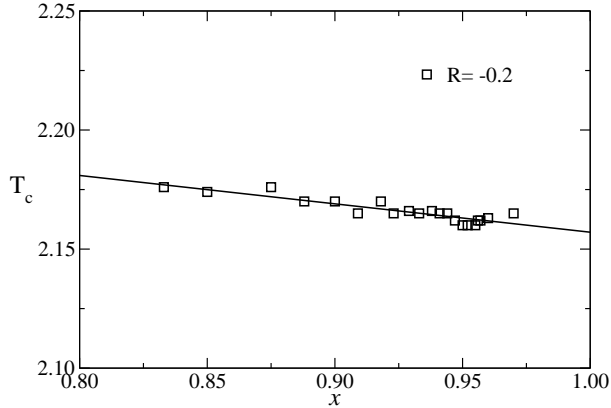}
\caption{ \label{fig:figure3}
Crossing temperatures for cumulants of systems of sizes $L$
and $L+2$ {\em versus} the ratio $x= L/(L+2)$.  Open squares are 
obtained from MC simulations with $R=-0.2$ and $H=2.0$. The straight 
line is a numerical fit to the data points.
}
\end{figure}
%%%%%%%%%%%%%%%%%%%%%%%%%%%%%%%%

To determine the phase diagram of the model, 
we use two different numerical methods, 
Monte Carlo (MC)~\cite{Cha86,Bin87,Lan00}
and Transfer Matrix (TM)~\cite{Nig82,Her83}.
Both methods have been used  
in statistical physics problems,
especially in Ising-type models. 
We are interested in the location of
the phase boundaries and the nature of the
transitions, whether they are of first- or
second-order, as well as if re-entrant is observed.  
Both methods are well-suited
to achieve those objectives.

In the present work, we use both methods
to determine the phase boundaries.
Even though MC is very
realible to ascertain the nature of the
phases,~\cite{Bin87,Lee90}
we elect to use the TM method
due mostly to its simplicity.
Once the phase boundaries are found 
by the TM method,
little further computational effort is needed 
to establish their nature~\cite{Nig82,Her83}.
We show results for the cases  $R= \pm 0.2, \pm 0.4,$ 
which, as we shall see, will provide the essential
features of the phase diagram.  We also consider 
the case $R=0$, 
which is known~\cite{Via09}, to check the reliability 
of our calculations.

In our MC calculations, we use the single-flip Metropolis 
algorithm~\cite{Met53} in square lattices of $L \times L$ spins, 
$ 8 \le L \le 128 $, with periodic boundary conditions.
We divide the lattice into two sub-lattices
$A$ and $B$, such that $A$ ($B$) is the set of
rows labeled with even (odd) indices.  We use
even values of $L$ to avoid frustration effects
at the edges of the $y$-direction, along which 
there is AF ordering in the SAF phase.

First, for a given set of the energy parameters and
temperature, we let the system equilibrate
after $10^7$ Monte Carlo steps (MCS).  Then
we collect the data for each additional
configuration generated by a sweep through
the lattice.  The data are stored in $10^3$ bins, 
each holding up to $10^4$--$10^5$ sets of
data points.  This will ensure that the 
autocorrelation time does not exceed the
bin size.
The average values in each bin
are used to determine the statistical averages and
the standard errors.  The corresponding error
bars are always smaller than the symbols we use
in all the graphs that follow.

In addition to the internal energy, specific heat, magnetization, 
and susceptibility.  We also calculate the SAF magnetization 
fourth-order cumulant, defined by:
 \begin{equation}
U_L = 1 - \frac{<M_s^4>_L}{3<M_s^2>_{L}^{2}}.
\label{cumulant}
\end{equation}
The quantities $<M_s^2>_L$ and $<M_s^4>_L$ 
are the second- and fourth-order moments of the
SAF magnetization, 
$<M_s> = \frac{1}{2}<(m_A - m_B)>$.
The quantities $<m_A>$ and $<m_B>$ are the sub-lattice 
magnetizations, 
with $<m_p> = <\frac{2}{L^2}\sum_{i\epsilon p} S_i^z> $ and
$p= A, B$.  
One of the properties of the fourth-order cumulant,
Eq.~2, is that as $T \rightarrow 0$, $U_L \rightarrow \frac{2}{3}$,
regardless the value of $L$.  At criticality, $U_L \rightarrow U^*$
in the thermodynamic limit~\cite{Bin81,Bin87,Ber88,Pec91}.

%%%%%%%%%%      Figure 4      %%%%%%%%%%%%%%
\begin{figure}
\includegraphics[width=8.0cm,angle=0]{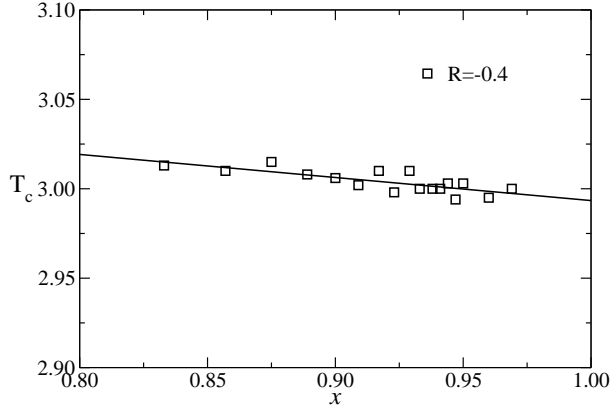}
\caption{ \label{fig:figure4}
Crossing temperatures for the cumulants of a system 
of size $L$, with another for size $L+2$, against
the ratio $x$.  
Open squares are from MC simulations with 
$R=-0.4$ and $H=2.0$, and
the straight line is a numerical fit.
}
\end{figure}
%%%%%%%%%%%%%%%%%%%%%%%%%%%%%%%%

The critical temperature is determined by the
intersections of the $U_L$ curves for systems of different sizes.
As an example, in Fig.~2 we plot the fourth-order 
cumulant {\em versus} temperature for the
cases $R=-0.2$, $H_c=2.0$, with system sizes 
$L= 8, 16, \dots, 128$.
The curves intersect nearly at the same point.  
In order to determine the critical temperature 
at the thermodynamic limit, in Fig.~3 we plot the 
crossing temperatures for two systems of linear sizes 
$L$ and $L=L+2$ {\em versus} the ratio $x= L/(L+2)$, 
with the same parameters as in Fig.~2.
Note that we use a finer scale for $T$,
as compared to the one used in Fig.~2.
The open squares are the crossing temperatures.
The straight line is a numerical fit to those
points, $T_c = 2.276 - 0.119\,x $.
The extrapolated value at $x=1$, the thermodynamic
limit, gives the critical temperature 
$T_c= 2.16 \pm 0.01$.
In Fig.~4 we show the crossing temperatures {\em versus} $x$
for $R=-0.4$, and the same parameters in Fig.~3.
Here, the straight line $T_c = 3.122 - 0.129\,x$ fits
the data.  In this case, after extrapolation we obtain
the thermodynamic value
$T_c = 2.99 \pm 0.01$.  As can be seen from
these figures, the temperature crossings
converge fairly rapidly to the thermodynamic
value of $T_c$.  That value can be inferred even when
very small lattices are used.  We employ this
procedure to obtain the critical lines in the $H$--$T$
space. 
Numerically, it becomes prohibitive time-wise 
to analyze the region $T < 0.2$, since it
becomes very difficult to obtain reliable
statistics.
Hence, in our MC simulations, we only 
treat cases $T \ge 0.2$.  

%%%%%%%%%%      Figure 5      %%%%%%%%%%%%%%
\begin{figure}
\includegraphics[width=8.0cm,angle=0]{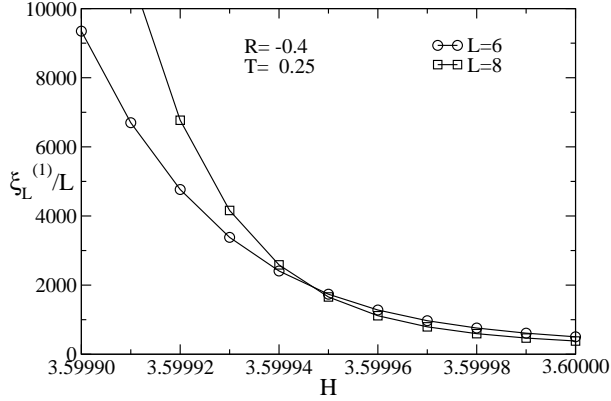}
\caption{ \label{fig:figure5}
First correlation length {\em versus} magnetic field for
infinite-length strips of widths $L=6, 8$ lattice spacings
obtained from Transfer Matrix method,
for the case $R=-0.4$ and $T = 0.25$.  
The curves cross at $H \simeq 3.599945$,
indicating a phase transition from 
the SAF phase to an induced
paramagnetic phase.
}
\end{figure}
%%%%%%%%%%%%%%%%%%%%%%%%%%%%%%%

At $T=0$, however, the model is trivially
solvable, so that we can determine the
critical temperatures and fields and thus
complete the phase diagrams to satisfaction.
There are two possible phases which,  depending
on the applied field, can be the ground-states
of the system: the SAF state, with its alternating
rows of up- and down- spins, and the induced
ferromagnetic (F) state.  At sufficient low fields $H$
the SAF state prevails, whereas at very large $H$
all the down-spins are flipped in the direction  
the field, hence the F state.  All other phases,
like the AFM-checkerboard or more exotic orderings,
will have higher energies than those of the SAF 
and F states, therefore they can be disregarded.
The ground-state energies of the SAF and F states 
are readily calculated, with results
$$
E_{SAF} = - 2(1- R)L^2,
$$
\begin{equation}
E_F = - (2R + H)L^2.
\end{equation}
By equating these energies, we determine
\begin{equation} 
H_c = 2 - 4R, 
\label{H_c}
\end{equation}
which is the field strength necessary to align
all the spins with the magnetic field without
expenditure of energy.

%%%%%%%%%%      Figure 6   %%%%%%%%%%%%%%

\begin{figure}
\includegraphics[width=8.0cm,angle=0]{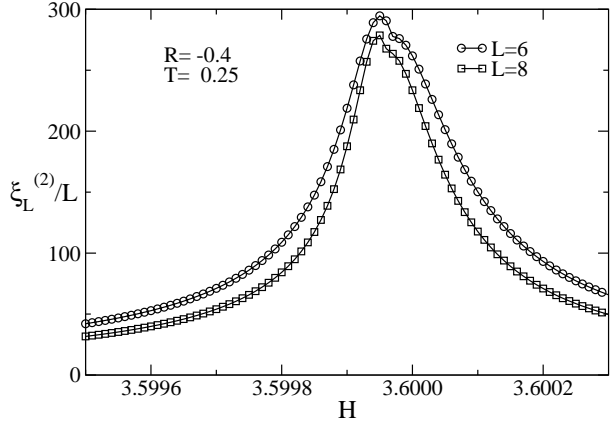}
\caption{ \label{fig:figure6}
Second correlation length {\em versus}
magnetic field for the same parameters as
in Fig.~5, as obtained from TM method. 
The absence of crossing indicates 
that the transition is of second order.
}
\end{figure}

%%%%%%%%%%%%%%%%%%%%%%%%%%%%%%%

We now proceed to the determination of the
phase-diagram of the system by using the
TM method~\cite{Nig82}.  
In addition to the location of the critical
temperatures and fields, the
method provides a simple criterion 
to establish the nature of the transition,
whether is of second- or first-order.  
It relies on two correlation lengths,
\begin{equation}
\xi_{L}^{(\alpha)} = \log^{-1}(E_0/E_{\alpha}),
\label{correlation}
\end{equation}
where $\alpha= 1$  denotes the first, and $\alpha= 2$ 
the second correlation
length. The quantities $E_0, E_1, E_2$ are the 
three largest transfer matrix  eigenvalues,
in descending order, for a strip of width $L$. 
The critical points are determined using two different lattice sizes
($L$, $M$), using 
\begin{equation}
L^{-1}\xi_{L}^{(\alpha)}(H,T) = M^{-1}\xi_{M}^{(\alpha)}(H,T). 
\label{renorm}
\end{equation}
We calculate the correlation lengths for infinite strips 
of widths $L= 2, 4, \dots, $ and $16$ 
lattice spacings, with periodic boundary conditions.  
The final results are 
extrapolated to $L \rightarrow \infty$.

%%%%%%%%%%      Figure 7    %%%%%%%%%%%%%%

\begin{figure}
\includegraphics[width=8.0cm,angle=0]{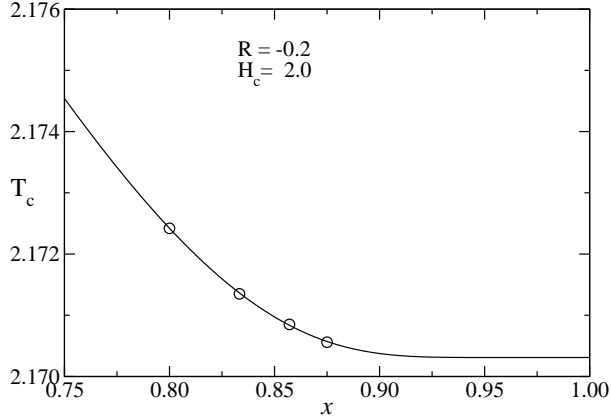}
\caption{ \label{fig:figure7}
Critical temperature as a function of the ratio $x$, between
the widths $L$ and $L+2$ of infinite-strips. 
The open circles are from TM results for $R= -0.4$ and 
$H_c = 2.0$, whereas
the solid line is a numerical fit.
}
\end{figure}

%%%%%%%%%%%%%%%%%%%%%%%%%%%%%%%

In Fig.~5, we plot the correlation
lengths $\xi_{L}^{(1)}$, 
for two infinite strips of widths $L= 6$ and $8$
with $T_c= 0.25$ and $R=-0.4$.
The crossing of the two curves determines
the critical field at $H_c = 3.599945$.
We use a similar plot with the second correlation
length $\xi_{L}^{(2)}$,
to unravel the nature of
the transition. 
Figure~6 shows the
second correlation length
for strips of widths $L=6$ and $8$.
The curves never cross, thus indicating
that the transition is of second-order.
We have examined the phase diagram
with this procedure throughout, and
conclude that the transitions
are always of second-order for the
entire range of parameters, and no 
re-entrant behavior is ever observed.

In order to obtain the thermodynamic
values of the critical temperatures and fields, 
 in Fig.~7 we plot the critical temperatures
$T_c$ against the ratio $x = L/M$, $M=L+2$.
We choose the same energy parameters
as those that were presented in Figs.~3 and 4,
to compare with the MC results. 
The open circles are the numerical 
results calculated from Eq.~4 for $R= - 0.2$ 
and critical field $H_c = 2.0$. 
The solid line is a nonlinear fit using 
$T_c = T_{\infty} + a \exp(-b/(1-x))$. 
Here, $T_{\infty}= 2.1703$, $a = 0.06827$, 
and $b = 0.6952$ are the numerically 
fitted parameters.  
The quantity $T_{\infty}$ is the extrapolated 
value for $T_c$ in the limit of infinite-width strips. 
The relative error between $T_c$ and $T_{\infty}$ 
for the largest width ratio used $x = (L/M) = (14/16)$ 
is about $0.01\,\%$, and less than $0.1\,\%$ for 
the smallest  ratio, $x= (8/10)$. 
We repeat the above procedure for
$R=-0.4$ and $H_c=2.0$, and the
results are displayed in Fig.~8.
Again, the open circles are obtained 
from Eq.~4.  The solid curve is given
by the nonlinear fit
$T_c = 3.00887 + 0.07571 \exp(-1.03606/(1-x))$. 
The extrapolated value for $T_c$ at $x=1$ gives 
the critical temperature of infinite-width strips
$T_{\infty}= 3.00887$. 
The relative error between $T_c$ and $T_{\infty}$ 
for the largest widths ratio $x = (10/12)$ is less 
than $0.005\%$ and about $0.1\%$ for 
the smallest ratio of strip widths, 
$x=(4/6)$. 
The other points of the phase-diagrams
can be calculated in a similar fashion.
Moreover, even 
for the smallest ratio, the estimated 
value for $T_c$ is already close to 
the extrapolated value of the 
infinite lattice.
One should also note the close
numerical agreement between 
$T_\infty$ found by the TM
method here with
the critical temperatures obtained
from the MC simulations of Figs.~3 and 4.
As will be shown in the following,
there is very good agreement between 
the results of TM and MC in all the
phase-diagrams.

%%%%%%%%%%      Figure 8      %%%%%%%%%%%%%%

\begin{figure}
\includegraphics[width=8.0cm,angle=0]{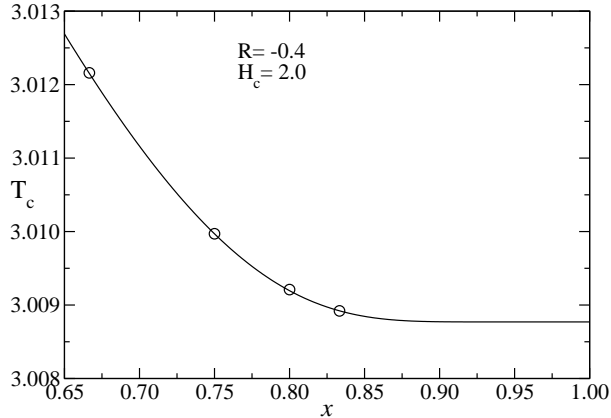}
\caption{ \label{fig:figure8}
Critical temperature {\em vs} $x$ for $R= - 0.4$ and $H_c=2.0$. The
open circles are TM results, and the solid line is a numerical
fit.
}
\end{figure}

%%%%%%%%%%%%%%%%%%%%%%%%%%%%%%%

The results from MC  and TM methods
are shown in Figs.~9 and 10, which depict the critical lines
for $R=0, \pm 0.2, \pm 0.4$.
The error bars are much smaller than the symbols in the
figures and are not shown in the graphs.
The critical lines obtained by the 
two methods show very good quantitative agreement
with each other,  and also they reproduce the known
result~\cite{Via09} for $R=0$.  Data for $R=0$ are shown in
the graphs to aid in the visualization of the effects 
of F ($R >0$) and AF ($R<0$)
$nnn$ couplings on the system.

Consider first the critical lines for $R = - 0.2$ and $- 0.4$, 
in Fig.~9.
The main effect of the {\em nnn} AF interactions is the expansion of
the SAF region in the phase diagram.  Such interactions
strengthen the SAF order. Thus it takes larger fields 
and/or temperatures to break this order.

This is to be contrasted with the case of a simple Ising  
model with {\em nn} AF interactions.  There,  ferromagnetic
{\em nnn} interactions reinforce the AF checkerboard
order and produce first-order transitions.
In our case, reinforcement of the SAF order
by {\em nnn} AF interactions does not produce 
first-order transitions.

%%%%%%%%%%      Figure 9      %%%%%%%%%%%%%%

\begin{figure}
\includegraphics[width=8.0cm,angle=0]{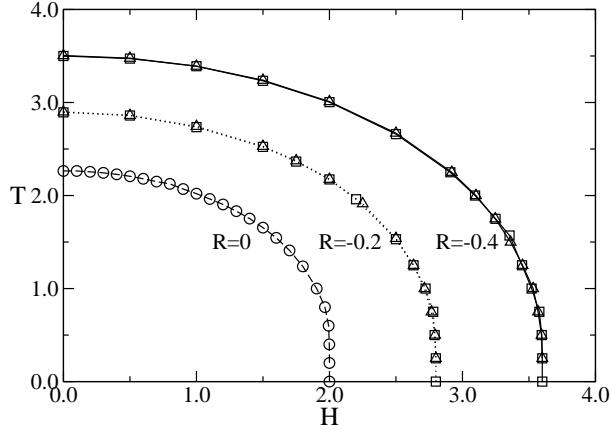}
\caption{ \label{fig:figure9} Critical lines of the
model for $R=0$, $-0.2$, and $-0.4$. 
Dotted lines connect the data points
for $R=-0.2$; full lines for $R=-0.4$; and dashed lines  for $R=0$.
Squares and circles were obtained from  MC  and triangles
from TM, except for the points at $T=0$, which were obtained 
from Eq.~4. 
}
\end{figure}

%%%%%%%%%%%%%%%%%%%%%%%%%%%%%%%

The critical lines for  $R = 0.2$ and $0.4$ are shown in Fig.~10.
There is a shrinkage of the region occupied by the SAF phase
as $R$ increases.    That is a result of the {\em nnn} interactions
competing with the local AF couplings, thus
weakening the SAF phase.   Hence, smaller fields and
temperatures are able to destroy the order.
The SAF phase region disappears altogether as $R \rightarrow \frac{1}{2}$,
which follows from setting $H_c = 0$ in Eq.~4.

%%%%%%%%%%      Figure 10      %%%%%%%%%%%%%%
\begin{figure}
\includegraphics[width=8.0cm,angle=0]{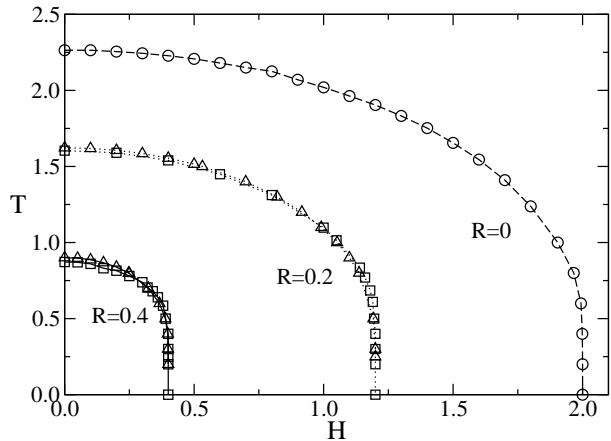}
\caption{ \label{fig:figure10}
Critical lines  for $R= 0$, $0.2$, and $0.4$.
The {\em nnn} couplings are now ferromagnetic 
and compete with the AF interactions, making it
easier to suppress the SAF phase.
The symbols have the same meaning as those in Fig.~9.
}
\end{figure}
%%%%%%%%%%%%%%%%%%%%%%%%%%%%%%%

To summarize, we studied the phase transitions of the SAF Ising model 
in  a uniform external magnetic field with {\em nnn} couplings
on a square lattice.
We used two numerical methods, Monte Carlo (MC) and Transfer Matrix (TM)
to obtain the critical lines in the ($H$--$T$) plane. 
We find that all transitions are of second-order and no evidence for
re-entrant behavior was observed.
Our main results are shown in Figs.~9 and 10. 
The critical properties of the model are marked 
by a transition line separating the SAF phase at low temperatures and fields 
from a paramagnetic phase at high temperatures and fields. 
The SAF order is reinforced when $R<0$, and 
depressed when $R > 0$, up until the limiting value 
$R_c = \frac{1}{2}$, at which the phase disappears entirely.

We thank FAPERJ (Brazilian agency) for financial support.
O.F.A.B. acknowledges support from the Murdoch
College of Science Research Program and a grant from the 
Research Corporation through the Cottrell College Science 
Award No.~CC5737.

%\begin{thebibliography}{99}

\end{document}